\documentclass[final]{aipproc}
\usepackage{amsmath,amssymb,amsfonts}
\usepackage{slashed}

\newcommand{\abs}[1]{\left| #1 \right|}

\newcommand{\psibar}{\overline{\psi}}
\newcommand{\tbar}{\overline{\theta}}
\newcommand{\Dbar}{\overline{D}}

\newcommand{\STr}{\mathrm{STr}}

\layoutstyle{6x9}

\begin{document}

\title{ 	The Phase Diagram for Wess-Zumino Models
}


\author{Franziska Synatschke\footnote{Speaker}}{
  address={Theoretisch-Physikalisches Institut,\,\,Friedrich-Schiller-Universit{\"a}t
Jena,
Max-Wien-Platz 1, D-07743 Jena, Germany}
}

\author{Holger Gies}{
  address={Theoretisch-Physikalisches Institut,\,\,Friedrich-Schiller-Universit{\"a}t
Jena,
Max-Wien-Platz 1, D-07743 Jena, Germany}
}

\author{Andreas Wipf}{
  address={Theoretisch-Physikalisches Institut,\,\,Friedrich-Schiller-Universit{\"a}t
Jena,
Max-Wien-Platz 1, D-07743 Jena, Germany} 
}

\begin{abstract} 
 Dynamical supersymmetry breaking is an important issue for applications of
 supersymmetry in particle physics. The functional renormalization group
 equations allow for a nonperturbative approach that leaves supersymmetry
 intact. Therefore they offer a promising tool to investigate
 dynamical supersymmetry breaking. Here we will employ this method to derive the
 phase diagram and a surprisingly rich RG fixed-point structure with 
 corresponding critical exponents for the $\mathcal N=1$ Wess-Zumino model  in
 two dimensions.
\end{abstract}

\maketitle

Supersymmetry (susy) is  an important ingredient for most theoretical
developments in quantum field theory beyond the standard model, supergravity
and string theory. The required susy breaking is believed to be of a
non-perturbative origin and therefore needs to be treated with non-perturbative
methods. Many approaches to investigate this problem, e.g. lattice calculations
 \cite{Feo:2004kx,Giedt:2006pd,Bergner:2007pu,Kastner:2008zc},  break susy
 explicitly, and it is hard to distinguish between dynamical and explicit susy
 breaking.
Functional renormalization group equations have proven to be a  successful
non-perturbative  method in quantum field theoretical calculations
\cite{Berges:2000ew,Pawlowski:2005xe}.
It is possible to formulate them in
superspace, guaranteeing to leave supersymmetry
intact
\cite{Rosten:2008ih,Synatschke:2008pv}. They offer a promising tool to
investigate dynamical susy breaking.
Here we consider the two dimensional Wess-Zumino model with one supercharge as
a model for dynamical susy breaking \cite{Witten:1982df}. These
proceedings follow our previous works on this subject
\cite{Gies:2009az,Synatschke:2009nm}.
 The action in superspace is given by
\begin{align}
	S=&\int d^2x \left.\left(-
	\frac12\Dbar\Phi\gamma_\ast D\Phi
	+W(\Phi)\right)\right|_{\tbar\gamma_\ast\theta}
\end{align}
with a real superfield $\Phi(x,\theta)=\phi(x)+\tbar\gamma_\ast\psi(x)
		+\frac12(\tbar\gamma_\ast\theta)F(x)$. The 
(super)covariant derivatives are given by 
$D={\partial}/{\partial\tbar}+i(\gamma^\mu\theta)\partial_\mu$ and
$\Dbar={\partial}/{\partial\theta}+i(\tbar\gamma^\mu)\partial_\mu$.
We use a chiral representation for the $\gamma$-matrices.
The susy transformations of the field components are 
	$\delta\phi=\varepsilon\gamma_\ast\psi,\;
	\delta\psi=(F+i\gamma_\ast\slashed{\partial}\phi)\varepsilon,\;\delta
F=i\overline{\varepsilon}\slashed{\partial}\psi,\;
	\delta\psibar=\overline\varepsilon(F-i\slashed{\partial}
\phi\gamma_\ast).$
After eliminating the auxiliary field using its EQM,  the on-shell
action
contains
one real bosonic field $\phi$ and one Majorana 
fermion field $\psi$, a  bosonic potential 
 $V(\phi)=W'(\phi)^2/2$ and a Yukawa interaction.
Unbroken susy ist characterized by a vanishing ground state energy.
It depends on the superpotential whether spontaneous susy
breaking is possible or not. For $W(\phi)\sim \phi^{2n}$ 
 susy is always unbroken whereas for $W(\phi)\sim\phi^{2n+1}$ spontaneous
 susy breaking is 
possible. We will focus on the latter in the following.

% ===========================================

The functional renormalization group can be formulated as a flow equation for
the effective average action $\Gamma_k$. This is a scale dependent functional
that
interpolates between the classical action $S=\Gamma_{k=\Lambda}$ at the UV
cutoff $\Lambda$ and the full
quantum effective action $\Gamma=\Gamma_{k=0}$ that includes all quantum
fluctuations. $\Gamma_k$ is determinated by the Wetterich equation
\cite{Wetterich:1992yh}
\begin{align}
  \partial_k\Gamma_k=
 \frac12 \STr\left\{\left[\Gamma_k^{(2)}+ R_k\right]^{-1}\partial_k  R_k\right\}
\,.\label{flow3}
 \end{align}
$k$ is an infrared regulator scale,
$(\Gamma_k^{(2)})_{ab}=\frac{\overrightarrow{\delta}}{\delta\Psi_a}
\Gamma_k\frac {\overleftarrow{\delta}}{\delta\Psi_b}$ denotes  the second
functional derivative of $\Gamma_k$, the indices $a,b$ summarize the field
components (internal and Lorentz indices, spacetime or momentum coordinates). 
$\Psi^T=(\phi,F,\psi,\psibar)$ denotes the collection of fields, not a
superfield.
$R_k$ is an infrared regulator 
derived from a cutoff action quadratic in the fields. 
A supersymmetric cutoff action is given by
$ \Delta S_k =\frac12  \int d^2x\; \Phi R_k
\Phi|_{\tbar\gamma_\ast\theta}.$ A general  supersymmetric regulator is a
function of covariant derivatives
	$R_k\equiv f(\Dbar\gamma_\ast D)\equiv r_1(\partial_\mu\partial^\mu)
	+r_2(\partial_\mu\partial^\mu)\Dbar\gamma_\ast D$. This decomposition is
	always  possible	because of
	$\{D_\alpha,\Dbar_\beta\}=-2i(\gamma^\mu)_{\alpha\beta}\partial_\mu$ and $ D^2=\Dbar^2=0\, .$
 $r_1$ is a mass-like regulator,  $r_2$
 corresponds to a kinetic term.

A consistent approximation
scheme to solve Eq. \eqref{flow3} is given by the derivative expansion
which in the
supersymmetric case implies an expansion in covariant derivatives.
The truncation of such an expansion preserves susy. The lowest-order
truncation is 
$
  \Gamma_k[\phi,F,{\psibar},\psi]=\int d^2x
	\left.(-\frac12\Dbar\Phi\gamma_\ast
	D\Phi+W_k(\phi))\right|_{\tbar\gamma_\ast\theta}
.$
In the following we  calculate the renormalization group flow of the
effective superpotential $W_k$. A ($\phi$-independent) wave function
renormalization $Z_k$ can be implemented via $\Phi\to Z_k\Phi$ in the
kinetic term.

The flow equation for the superpotential is obtained by projecting Eq.
\eqref{flow3} onto the part linear in the auxiliary field
\begin{align}
	\partial_kW_k(\phi)=&\frac12\int \frac{d^2p}{4\pi^2}
	\frac{(r_2+1)\partial_kr_1-(r_1+W''_k(\phi ))\partial_kr_2}
	{(r_1+ W''_k(\phi ))^2
   +p^2 (r_2+1)^2}\label{eq:FlowEQ}.
\end{align}
	We are interested in superpotentials with spontaneous susy
breaking,
	therefore their highest power is odd, as is the highest power in the
second
	derivative. The regulator function $r_1$ amounts to just a
$\phi$-independent
	shift so that we can set it to zero without loss of generality. 
In the following we  choose 
$r_2=\left(\abs{{k}/{p}}-1\right)\theta(1-{p^2}/{k^2})$ for which the  momentum
integration can be performed analytically, and the flow equation simplifies to
$
	\partial_kW_k(\phi)
	=-({k}{W''_k(\phi)})/({4\pi}({k^2+W''_k(\phi)^2})).$
However, this simple regulator leads to artificial
divergences if a wave function renormalization is included and a
stronger regulator e.g.
\mbox{$r_2=\left({k^2}/{p^2}-1)\theta(1-{p^2}/{k^2}\right)$} must be
used.

\begin{figure}
\includegraphics[height=.25\textheight]{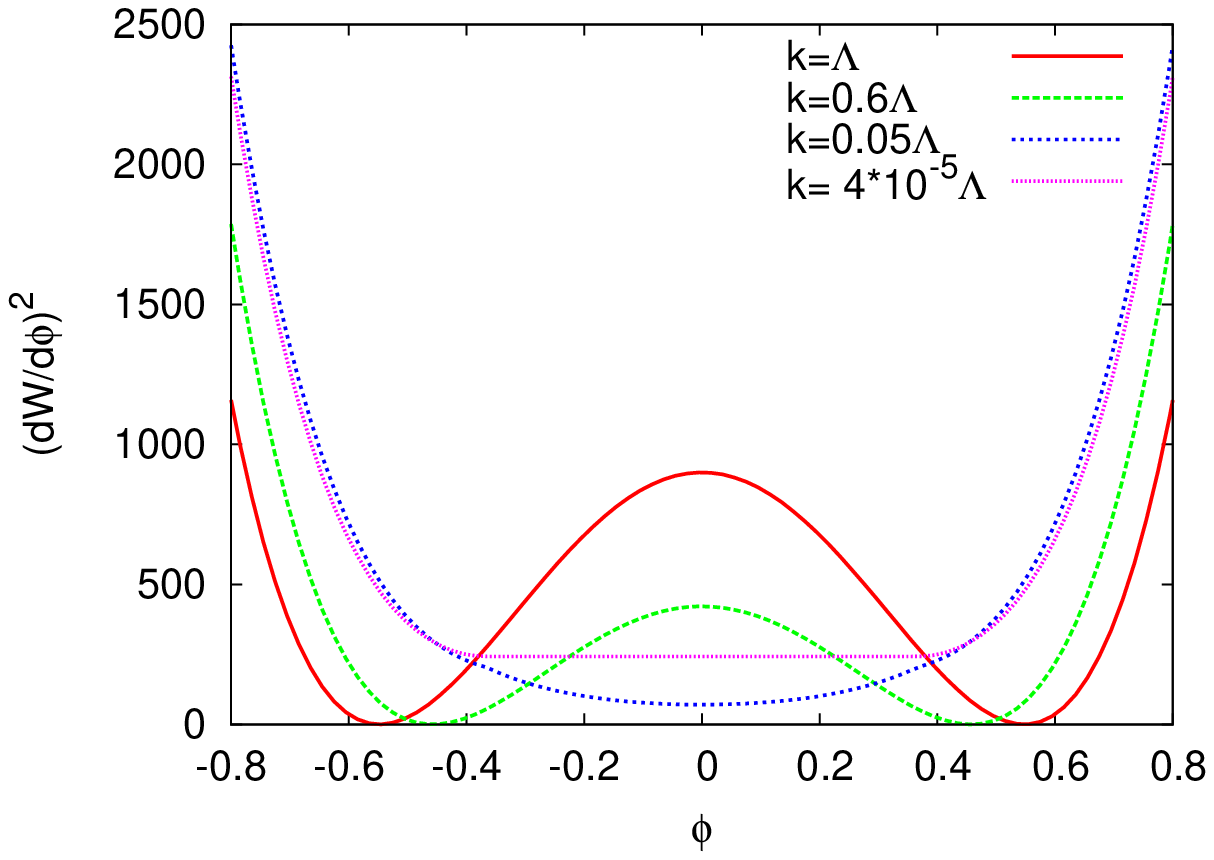}
\includegraphics[height=.25\textheight]{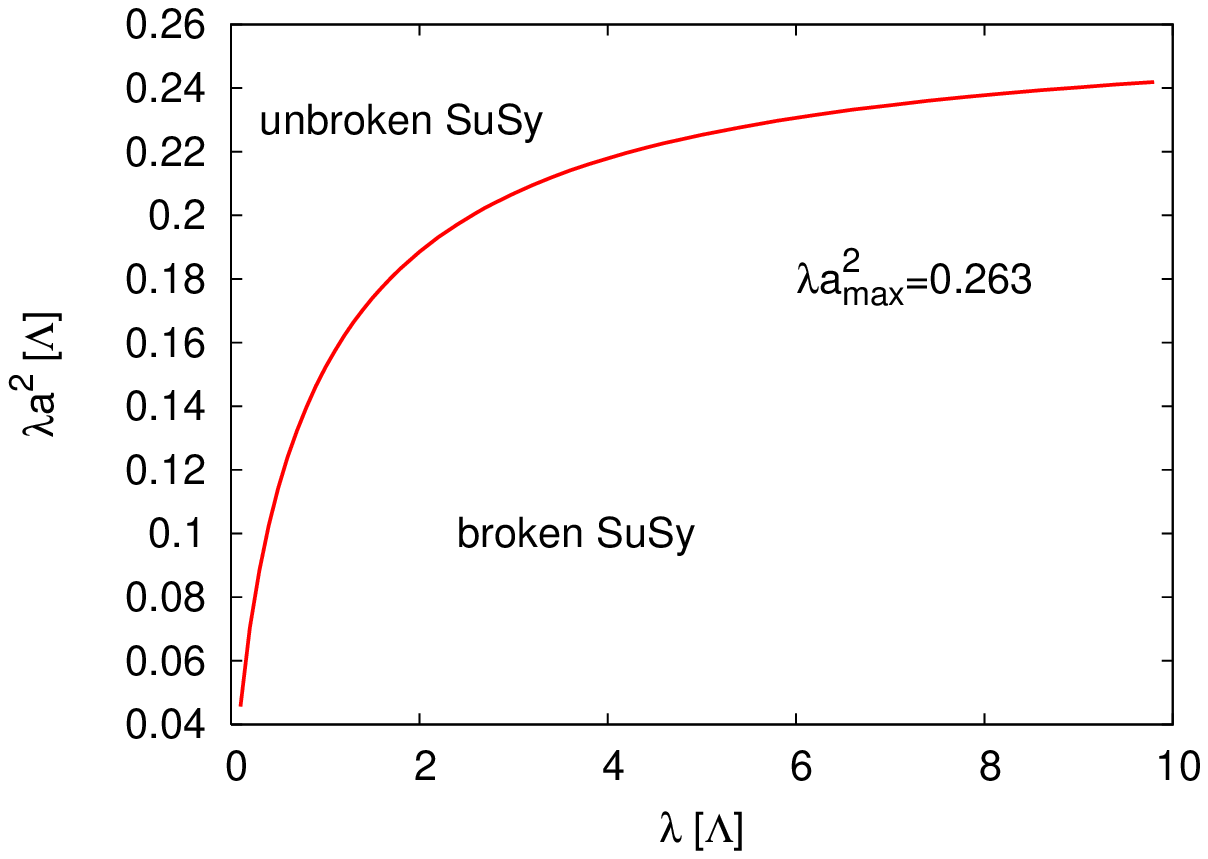}
\caption{\emph{left panel:} Flow of the effective potential $W'^2$ with
$\lambda=0.1\,\Lambda$ and $a^2=0.3$. 
\emph{right panel:} phase diagram \label{fig1}}
\end{figure}
It is possible to solve this partial differential equation numerically. Here we
employ a further approximation and expand the superpotential into a power
series, $W_k'=\lambda_k(\phi^2-a_k^2)+\sum_{n=2}b_{2n,k}\phi^{2n}$. The flow
equation then turns into a system of coupled ordinary differential equations
for the coefficients.  As initial conditions at $k=\Lambda$ we take
$b_{2n,\Lambda}=0$ and nonzero values for $\lambda_\Lambda$ and $a_\Lambda^2$.
The flow of the bosonic potential $V_k(\phi)= W_k'(\phi)^2/2$ with
$\lambda_\Lambda=0.1\Lambda$ and $a^2_\Lambda=0.3$ is shown in Fig. \ref{fig1} (left panel). At the cutoff
scale  we start with a double-well potential and unbroken susy
($E_0=0$). As the scale $k$ is lowered to the infrared, a single-well potential
emerges and we end up in the phase of broken susy. We find that in the
regime with broken susy the curvature of the bosonic potential at
the minimum and therefore the bosonic mass
goes to
zero with the RG scale $k$ as $
 m(k)\sim k^{1/\nu}$. This behavior is
governed by a critical exponent $\nu$ 
 which obeys the super-scaling relation  
\[\nu=\frac{2}{2-\eta},\quad \eta=-k\partial_k\ln Z_k^2,\quad\nu\simeq1.3,\]
where $\eta$ denotes the
anomalous dimension. We emphasize that any measurement (e.g.
lattice calculations) involves an IR cutoff (e.g. the lattice volume). Hence we
predict that any measurement will yield a bosonic mass proportional to the
measurement scale provided by this IR cutoff.

% ===========================================

The coupling $a^2_\Lambda$ is a control parameter for susy breaking. 
In Fig. \ref{fig1} (right panel) the phase diagram in the control-parameter
plane $(\lambda_\Lambda,a^2_\Lambda\lambda_\Lambda)$ is shown. As a signal for
susy breaking we use a nonvanishing ground state energy. We find  a
maximal value for susy breaking at $\lambda_\Lambda
a^2_\Lambda\simeq 0.263$. This agrees with a
qualitative argument given by Witten \cite{Witten:1982df} that spontaneous susy
breaking is not possible for large values of $a^2_\Lambda$.

\begin{figure}
\includegraphics[height=.25\textheight]{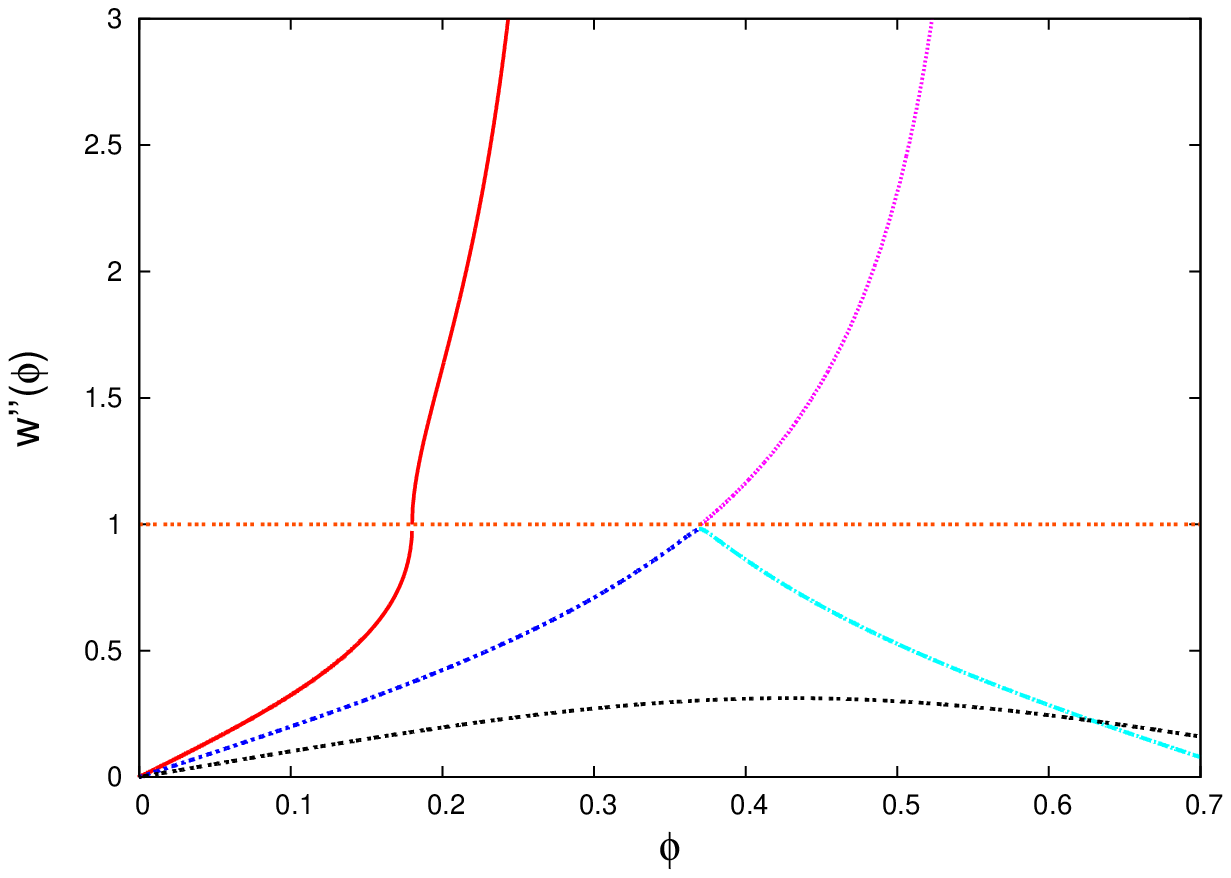}
\includegraphics[height=.25\textheight]{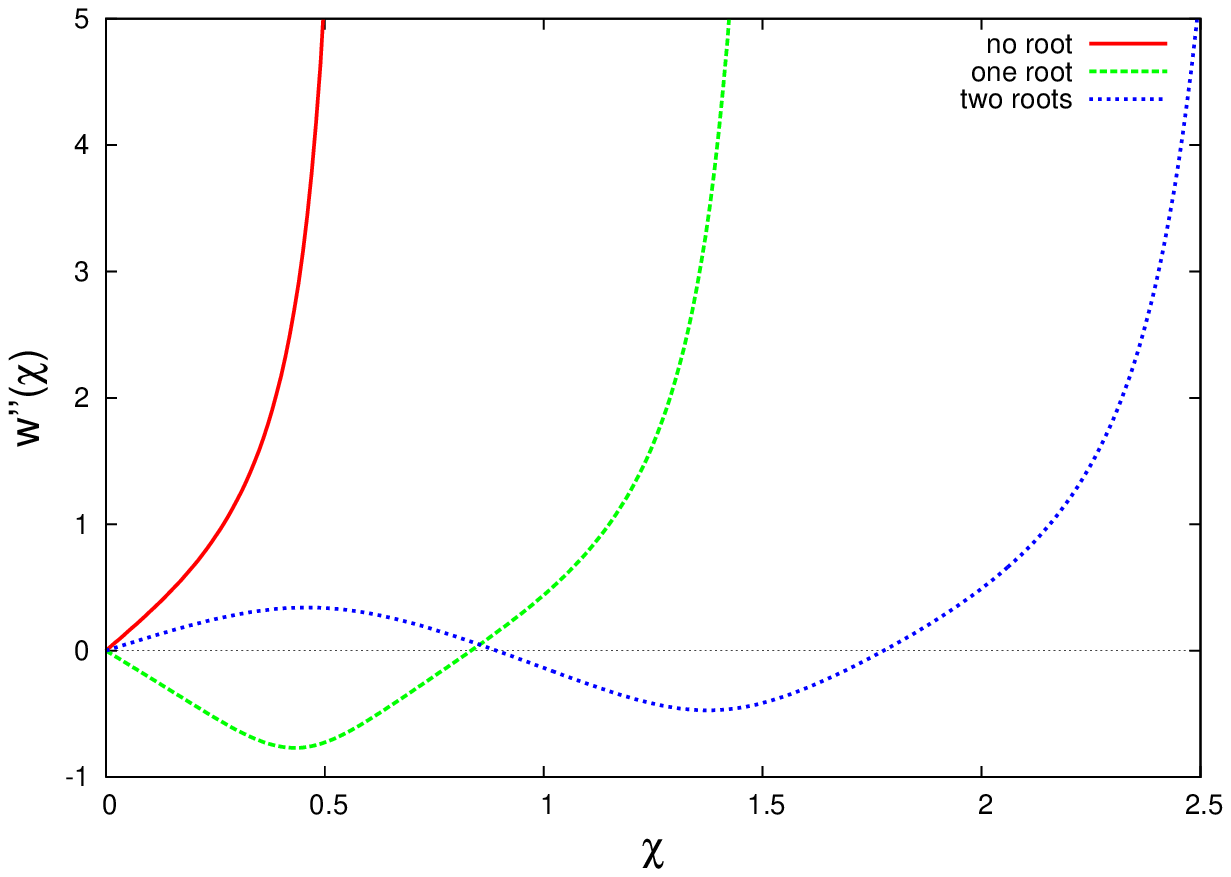}
\caption{\emph{left panel:} fixed point solutions in the LPA-approximations;
\emph{right panel:} fixed point solutions in the next-to-leading order
approximation\label{fig:FPSolutions}.}
\end{figure}
Now we investigate the fixed points. For this we
have to rescale 
the flow equation for the superpotential to dimensionless quantities
$w_k(\phi)=W_k(\phi)/k$ and $t=\ln(k/\Lambda)$.
	The  field $\phi$ in two dimensions is dimensionless. 
	The fixed points are characterized by the condition
$\partial_t w_\ast=0$. This leads to a nonlinear ordinary differential
equation with a singularity at $w_k''(\phi)=1$. 
The superpotential has two relevant directions corresponding to the coefficients
of the terms $\phi^0$ and $\phi^1$.
As	 only the second derivative of the superpotential enters on the right
hand side of Eq. \eqref{eq:FlowEQ} it is sufficient to consider the
second derivative of the fixed-point equation to get rid of the  IR-unstable
directions.
To leading order of the derivative expansion (local potential approximation)
we find a continuum of oscillating solutions and solutions confined to a finite
interval (see Fig. \ref{fig:FPSolutions} left panel). 
At next-to-leading order where a field independent wave function
renormalization is included  we find a discrete number of
fixed point
 potentials that for
 large fields behave as $w''\sim\chi^{2/\eta-1}$ with $\chi=Z_k\phi$. They are
characterized by the
 number of nodes (see Fig. \ref{fig:FPSolutions} right panel). For the first
 solutions we find the anomalous dimensions and critical exponents
 $\eta=0.4386$, $\nu=1.2809$ (no nodes), $\eta=0.20$, $\nu=1.11$ (one node)
 and $\eta=0.12$, $\nu=1.06$ (two nodes).

To conclude, as already demonstrated in supersymmetric quantum mechanics the
formulation in
superspace is suitable to extend the functional renormalization group to
supersymmetric theories. We were able to derive the phase diagram for
susy breaking and to determine the fixed point structure in the local
potential approximation. 
So far we have only considered a constant wave function renormalization. To
go further in the derivative expansion we have to consider field dependent and
probably momentum dependent wave function renormalization.

Helpful discussions with G.~Bergner, C.~Wozar, T.~Fischbacher
  and T.~Kaestner are gratefully acknowledged.  This work has been supported
  by the Studienstiftung des deutschen Volkes and the DFG under GRK 1523,
  Wi 777/10-1, FOR 723 and \mbox{Gi 328/5-1}.

\enlargethispage{6\baselineskip}


\begin{thebibliography}{11}
\expandafter\ifx\csname natexlab\endcsname\relax\def\natexlab#1{#1}\fi
\providecommand{\enquote}[1]{``#1''}
\expandafter\ifx\csname url\endcsname\relax
  \def\url#1{\texttt{#1}}\fi
\expandafter\ifx\csname urlprefix\endcsname\relax\def\urlprefix{URL }\fi
\providecommand{\eprint}[2][]{\url{#2}}

\bibitem[Feo(2004)]{Feo:2004kx}
A.~Feo, \emph{Mod. Phys. Lett.} \textbf{A19}, 2387--2402 (2004),
  \eprint{hep-lat/0410012}. %%CITATION = HEP-LAT/0410012;%%

\bibitem[Giedt(2006)]{Giedt:2006pd}
J.~Giedt, \emph{Int. J. Mod. Phys.} \textbf{A21}, 3039--3094 (2006),
  \eprint{hep-lat/0602007}. %%CITATION = HEP-LAT/0602007;%%

\bibitem[Bergner et~al.(2008)]{Bergner:2007pu}
G.~Bergner, T.~Kaestner, S.~Uhlmann, and A.~Wipf, \emph{Annals Phys.}
  \textbf{323}, 946--988 (2008), \eprint{0705.2212}. %%CITATION = 0705.2212;%%

\bibitem[Kaestner et~al.(2008)]{Kastner:2008zc}
T.~Kaestner, G.~Bergner, S.~Uhlmann, A.~Wipf, and C.~Wozar, \emph{Phys. Rev.}
  \textbf{D78}, 095001 (2008), \eprint{0807.1905}. %%CITATION = 0807.1905;%%

\bibitem[Berges et~al.(2002)]{Berges:2000ew}
J.~Berges, N.~Tetradis, and C.~Wetterich, \emph{Phys. Rept.} \textbf{363},
  223--386 (2002), hep-ph/0005122. %%CITATION = HEP-PH/0005122;%%

\bibitem[Pawlowski(2007)]{Pawlowski:2005xe}
J.~M. Pawlowski, \emph{Annals Phys.} \textbf{322}, 2831--2915 (2007),
  \eprint{hep-th/0512261}. %%CITATION = HEP-TH/0512261;%%

\bibitem[Rosten(2008)]{Rosten:2008ih}
  O.~J.~Rosten,
  \emph{On the Renormalization of Theories of a Scalar Chiral Superfield},
   \eprint{0808.2150}
  %%CITATION = ARXIV:0808.2150;% 

\bibitem[Synatschke et~al.(2009{\natexlab{a}})]{Synatschke:2008pv}
F.~Synatschke, G.~Bergner, H.~Gies, and A.~Wipf, \emph{JHEP} \textbf{03}, 028
  (2009{\natexlab{a}}), \eprint{0809.4396}. %%CITATION = 0809.4396;%%


\bibitem[Witten(1982)]{Witten:1982df}
E.~Witten, \emph{Nucl. Phys.} \textbf{B202}, 253 (1982). 
%%CITATION = NUPHA,B202,253;%%

\bibitem[Gies et~al.(2009)]{Gies:2009az}
H.~Gies, F.~Synatschke, and A.~Wipf  (2009), \eprint{0906.5492}.
%%CITATION = 0906.5492;%%

\bibitem[Synatschke et~al.(2009{\natexlab{b}})]{Synatschke:2009nm}
F.~Synatschke, H.~Gies, and A.~Wipf  (2009{\natexlab{b}}), \eprint{0907.4229}.
%%CITATION = 0907.4229;%%

\bibitem[Wetterich(1993)]{Wetterich:1992yh} 
C.~Wetterich, \emph{Phys. Lett.} \textbf{B301}, 90--94 (1993). 
%%CITATION = PHLTA,B301,90;%%



\end{thebibliography}
\end{document}